\documentclass[a4paper, 10 pt, conference]{ieeeconf}  

\IEEEoverridecommandlockouts                              
\usepackage{graphicx}
\usepackage{amsmath,amsfonts,amssymb}

\title{\LARGE \bf
Reliability Evaluation of Phasor Measurement Unit Considering Failure of Hardware and Software Using Fuzzy Approach
}

\author{Evan Carollo and Zikai Xu} 

\begin{document}

\maketitle
\thispagestyle{empty}
\pagestyle{empty}

\begin{abstract}

The wide-area measurement system (WAMS) consists of the future power system, increasing geographical sprawl which is linked by the Phasor measurement unit(PMU). Thus, the failure of PMU will cause severe results, such as a blackout of the power system.  In this paper, the reliability model of PMU is considered both hardware and software,  where it gives a characteristic of correlated failure of hardware and software.  Markov process is applied to model PMU, and reliability parameters are given by using symmetrical triangular membership for Type-1 fuzzy reliability analysis. The paper gives insightful results revealing the effective approach for analyzing the reliability of PMU, under a circumstance which lack of sufficient field data.


\end{abstract}

\section{INTRODUCTION}

"The phenomenal development of modern power systems
in terms of both geographical sprawl and technological
innovations has emerged as one of the most complex engineering systems in existence. A modern power system involves
thousands of components for generation, transmission, and distribution before reaching the consumer. To consistently provide
power to residential and commercial customers around the
clock, the generation, transmission, and distribution systems
must have proper coordination. Therefore, it is very important
to periodically monitor the condition or “health” of the system.
A wide-area measurement system (WAMS) is a powerful tool
for monitoring the health of a power system. Phasor measurement units (PMUs) are key components of a WAMS, providing
precise and real-time grid measurements that are time-stamped
according to a common time reference. The synchronized phasor measurements can be used for improving the reliability
of the system."

"Reliability modeling of PMU has received ample research interest owed to its critical role in delivering WAMS services. Unfortunately these models are restricted in terms of possible failure mode of PMU. In fact, none of them accommodate hardware–software interaction failures in modeling PMU reliability, which has been an area of research interest, partly because of its recent identification and partly because of consequences that such failures have led to some of the contemporary safety critical systems. In recent years, however, there has been a lot of emphasis on identifying, modeling, and quantifying the effect of hardware– software interaction failures on system reliability in myriad cases including integrated circuit (IC) fabrication technologies, jet propulsion systems and so on. So far, all the PMU models developed have assumed the central processing unit (CPU) module as a combined hardware–software unit similar to a digital signal processor or a microprocessor on which software executes for achieving the intended functions. This assumption implies that hardware and software are independent and that the PMU system fails if the hardware or software subsystems fail alone. Based on this assumption, the software subsystem has been considered to be in series with the hardware subsystem from a reliability perspective [3]. This assumption imposes a serious limitation for systems, particularly for embedded systems, where hardware and software work in close proximity with each other. PMU is an embedded device, and thus its hardware and software subsystems bear close correlation for proper functioning. In fact, this interaction between hardware and software subsystems is present throughout the PMU system, not just in the CPU model. For example, high-precision time clock signal productions necessitate the inclusion of a high-precision crystal oscillator within the GPS module of the PMU [2]. The crystal oscillator can switch between track mode and replacement mode and this switching is a functional switching as opposed to a physical switching governed by software logic. It has to be noted that such software controlled functionalities have not been accounted for in the existing reliability models, since software subsystem has been accounted only as a part of the CPU module. Similarly, communication module functions based on the software protocol stacks implemented and software bugs therein, if any, have to account for is the overall PMU reliability model. This has encouraged us to analyse the reliability of PMUs considering the interactions between hardware and software."

"The reliability analysis techniques discussed in
[2]–[4] do not consider any data uncertainty in the reliability
parameters when evaluating the reliability of a PMU. Because
PMUs have been introduced only recently, statistical data on
the operation of PMUs is very sparse. The uncertainty in sparse
statistical data consequently leads to difficulty in estimating the
reliability parameters of PMU components. Therefore, rather
than attempting to estimate a single value, it is more appropriate
to estimate the range of a reliability parameter to account
for uncertainties. Because of the scarcity of statistical data,
practical knowledge and engineering decisions might be used
in estimating the ranges of the reliability parameters. Generally,
practical knowledge and engineering decisions are characterized by fuzzy linguistic descriptions. The advantage of fuzzy
logic is that the representation of knowledge is very explicit
using “IF–THEN” relations."

The main contribution of the paper is quantitative reliability
evaluation of PMU using type-1 fuzzy sets. Here, a state
Markov model has been taken into consideration which
includes the power supply module. Power supply module is an
important component as it supplies power to the PMU.  The reliability parameters are calculated using
the fuzzy set symmetrical triangular membership function. This
simplifies the calculations. Moreover, type-1 fuzzy set handles
the data uncertainty. Here, the data uncertainty has been
considered for all the modules for better analysis of PMU
reliability. 

\section{Phasor Measurement Unit }
\begin{figure}[t]
\centering
\includegraphics[width=0.5\textwidth]{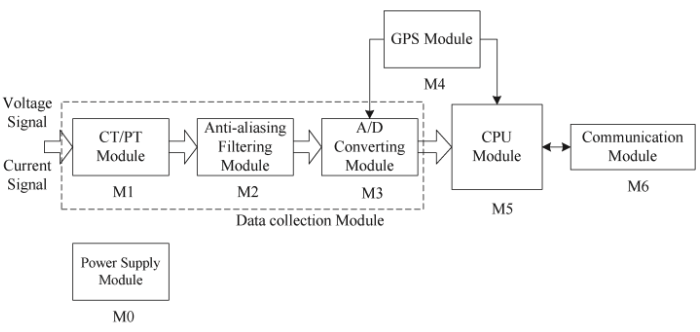}
\caption{}
\end{figure}
The block diagram of Phasor Measurement Unit (PMU) is given in the Fig.1. One can observe that all the modules of PMU are connected in a series where one module of failure will cause failure of the system. The current transformer/potential transformer(CT/PT) model(M1) convert the analogue three-phase voltages and currents to the safer level. In the next level, these signal are passed through Anti-aliasing filtering module(M2) to filter out high-frequency noises, which are subsequently converted to digital signals by A/D converting Module(M3). The CPU module (M5) obtain the magnitude, phase and power frequency of the signal processed by M3. The GPS module(M4) attach highly accurate time synchronisation signal to the measurements. Finally, Communication module (M6) transmit data to other terminals to further process PMU data. The power supply module (M0)  provides power for all modules in the system. 

Here one can observe that all the components of PMU models consider only hardware parts; however, the effects of software causing hardware failure or hardware  causing software failure are ignored. In fact, most of research assume that hardware and software are independent, and the only interaction between hardware and software are considered in CPU module. Although an embedded system such as PMU, each module are connection in series and all controls flow from one module to the next module; the real control behind this is software. The purpose of this project is to improve the accuracy of reliability model of PMU considered software and hardware interactions. This paper provides a unified software and hardware PMU model other than traditional PMU reliability analysis only focusing on either hardware or software.
\begin{figure*}[t]
  \includegraphics[width=2\columnwidth]{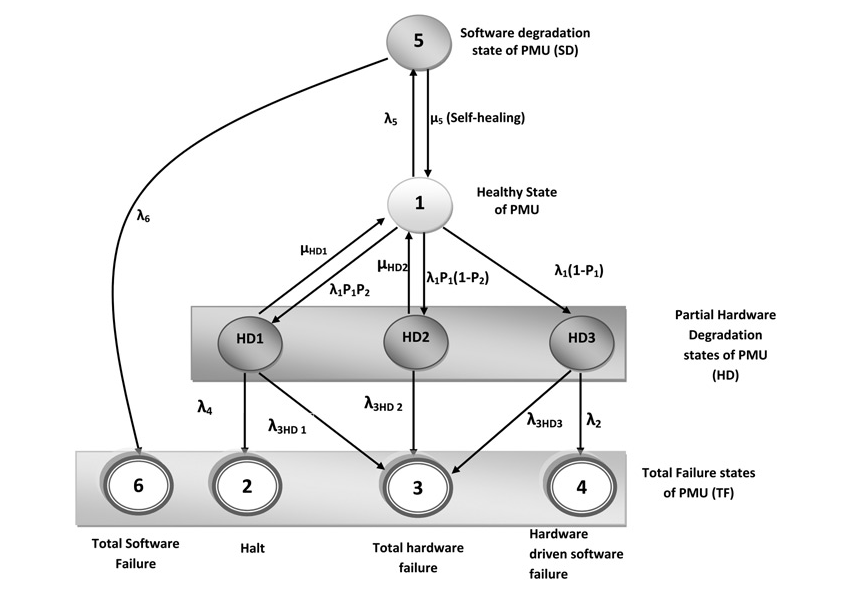}\caption{}
\end{figure*}
\section{Unified PMU model }
\subsection{ Introduction of Model}
The scheme of the Unified PMU model is given on Fig.2 where it demonstrate the Markov model and how hardware and software interact with each other.  In the unified PMU model, there are total 7 states. Each different state can depict hardware failure, software failure or hardware-software interaction failures in a PMU. Here, we explain each state's representations in our unified PMU model. State 1 stands for UP state of the PMU where PMU is health and working functionally. From state 1, the system can jump into either software degradation(SD) state or partial hardware degradation (HD) state. First, we introduce HD state. There are three possible HD states denoted as HD1, HD2 and HD3.
\begin{itemize}
    \item HD1: partial degradation may be detected but may not recover.
    \item HD2: degradation is detected and recovered by software.
    \item HD3: degradation is undetected and go faliure
\end{itemize}
Any HD state can further lead the total failure of the PMU system (TF), where the last rectangular is shown the TF. At HD3 state, undetected hardware degradation dose no given any signal to PMU system;  and transition from HD3 to state 4 represents the hardware-software interaction falures. State 5 represents the software degradation(SD). Errors in the SD may accumulate and finally cause the total failure of the system (TF).  
\subsection{Reliability Analysis of The Model}
The pre-assumption of the model is a Markov process, where the future state is independent on  past states. Thus, transition paths: 1 to HD1/HD2 to 3; 1 to HD3 to 4; 1 to SD to 6  are independent to each other. Thus, we can write the PMU reliability as 
\begin{equation}
    R_{PMU}(t) = R_{HW}(t)R_{SW}R_{HW\&SW}(t).
\end{equation}

Hardware reliability can be represented by Weibull model 
\begin{equation}
    R_{HW}(t) = e^{-\lambda t^{\beta}},
\end{equation}
where $\lambda$ is the failure rate of hardware components and $\beta$ is the shape parameter of the Weibull distribution function.

Software reliability can be definded by non-homogenous process model with mean time function $m(t)$. If the software startup time T, the software reliability is  \begin{equation}
    R_{SW}(t) = e^{(-m(t+T)-m(T))},
\end{equation}
where t is the testing time and $a$, $b$ are number of faults to be detected and fault detection rate, respectively. 

According [], we  can find the probability of hardware-software working at time $t$ as 
\begin{equation}
    R_{HW\&SW}(t)  = Q_{1}(t)+ Q_{HD1}(t)+ Q_{HD2}(t)+ Q_{HD3}(t),
\end{equation}
where $Q_{1}(t)$ is probability at state 1 at time {t}. Similarly,  $Q_{HD1}(t)$ is the probability at state HD1, $Q_{HD2}(t)$ is the probability at state HD2, $Q_{HD3}(t)$ is the probability at state HD3.
Therefore, we can express the overall system reliability as 

\begin{align}
     R_{PMU}(t) &=R_{HW}(t)R_{SW}R_{HW\&SW}(t)\\
      &= e^{-\lambda t^{\beta}} e^{(-m(t+T)-m(T))}\\
      & \times (Q_{1}(t)+ Q_{HD1}(t)+ Q_{HD2}(t)+ Q_{HD3}(t)).\label{interaction}
\end{align}
\begin{figure}[t]
\centering
\includegraphics[width=0.4\textwidth]{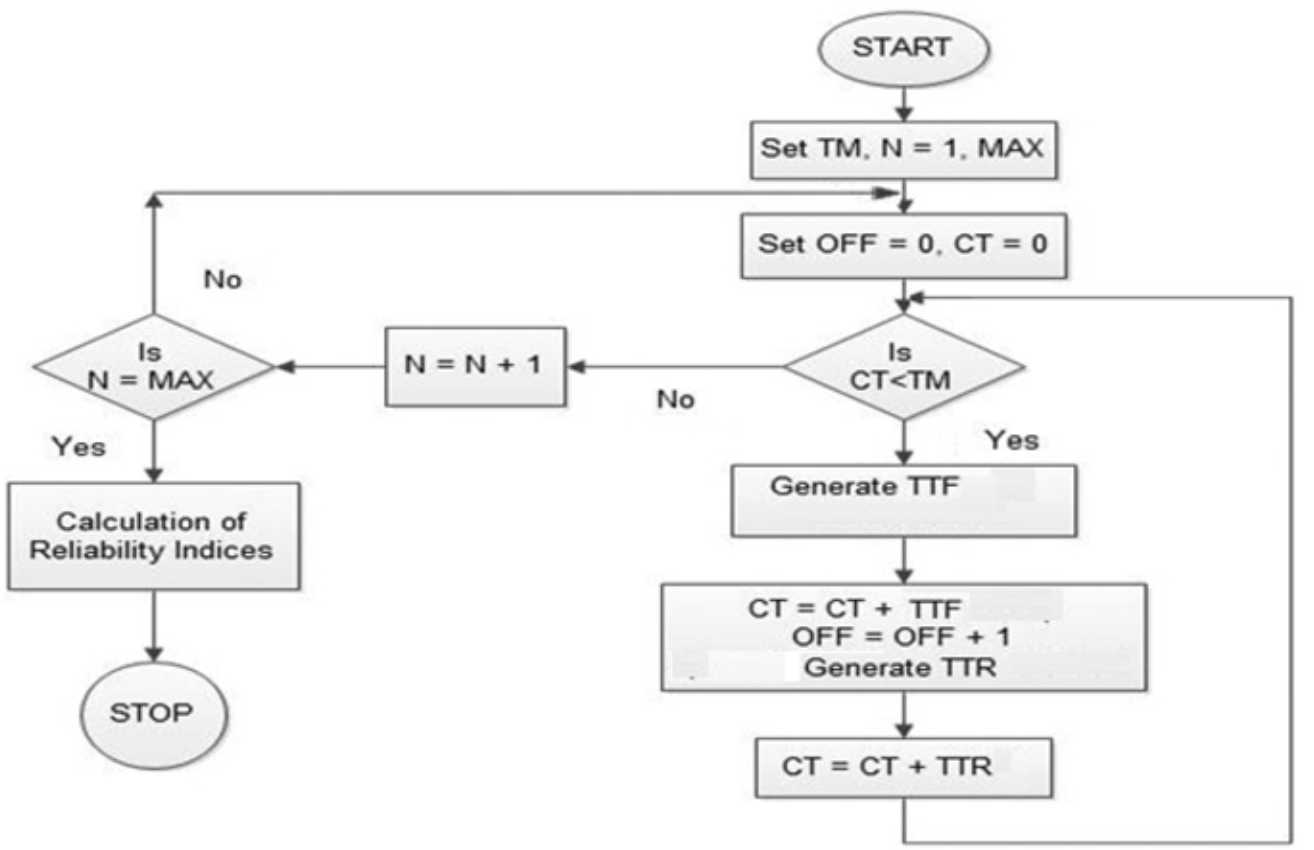}
\caption{}
\end{figure}
Thus, \eqref{interaction} gives the hardware-software interaction failures. Given PMU failure data, one can find hardware-software interaction failure; however, PMUs are launched recently and failure data are limited. Also, we cannot have accurate reliability index for PMUs. Thus, we use fuzzy approach to obtain reliability index of PMU, and then we apply Markov simulation to find failure data of PMU. 

\section{ Fuzzy  Approach }  
Reliability parameters are the key factor to  analyze  PMU; however, failure data  is not sufficient to help us to accurately analyze  PMU, causing uncertainty in reliability parameters of PMU components.  Thus, instead of giving one single value of reliability parameter, a range of  reliability parameter is much more reasonable and accurate results. Here, we use Type-1 fuzzy sets to handle the uncertainty of the data.  According to [], we can have the following steps to determine the Type-1 fuzzy reliability of PMU:
\begin{itemize}
    \item[(1)]  To determine the input data. Here, input data is that the failure and repairer rates are in a symmetrical triangular membership functions.
    \item[(2)]  $\alpha$-cut sets of input data are established for $\alpha=[0,1]$
    \item[(3)] The parameters of the equivalent reliability model for any $\alpha$ are calculated. 
    \item[(4)] $\alpha$-cut are used to calculate fuzzy outputs.  
\end{itemize}
The availability and unavailability of PMU is given in the Fig.3(a) and Fig.3(b); and Fig.3(c) and Fig.5(d) give a range of repair rate and failure rate of PMU.
\begin{figure}[t]
\centering
\includegraphics[width=0.5\textwidth]{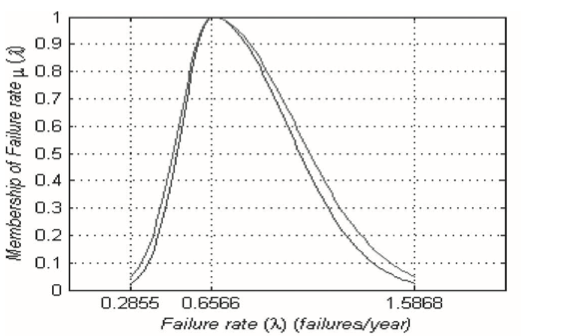}
\caption{}
\end{figure}
\begin{figure}[t]
\centering
\includegraphics[width=0.5\textwidth]{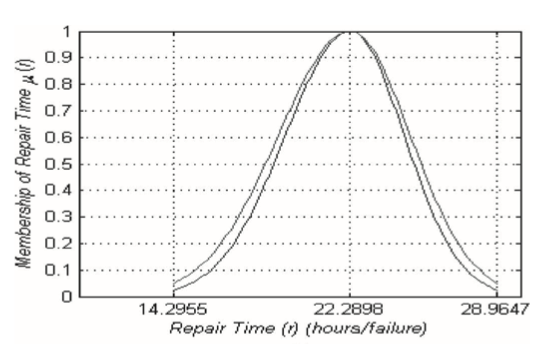}
\caption{}
\end{figure}
Finally, we choose failure rate as 0.6566 F/year and repair rate as 22.2898 Hs/year.   

\section{ PMU Failure Data from Simulation }
In this project, we apply Markov simulation to estimate failure data of PMU. First of all, based on failure rate and repair rate given in the previous section, we use simulation to generate random number for PMU time to failure and PMU time to repair. Then, based on the  time to failure and number of failure to find hardware and software interaction failure reliability.  
The detail steps of simulation as followings and the flow chart of the simulation is given on Fig.4.
\begin{itemize}
    \item[(1)] Determine the mission time, we set mission time (TM) 10 years. Total number of simulation as 10000.
    \item[(2)] Initialize number of simulation N as 0.
     \item[(3)]Number of failure and counter set to 0
     \item[(4)] Check whether counter is less than mission time. If condition is true, another simulation is applied until condition reach false. Use random number to generate failure time. 
    \item[(5)] Number of failure and failure time are accumulated. Repair time also calculate.
     \item[(6)] Add repair time to counter .
     \item[(7)] Repeat step 4,5,6 if counter<  mission time.
      \item[(8)] Based on the simulation results to obtain reliability indices. 
\end{itemize}

\section{ Results and Analysis}

This section gives results of PMU reliability analysis based on the model proposed before. 
\subsection{Fitting the PMU failure data to the model}

As we discussed in the previous section, we have 6 parts in the system. That the path  from HD3 to 4 is undetected hardware failures leading hardware related software failures. Assume that  all failure rate are considered as 0 except $\lambda_{1}$ and $\lambda_{2}$, we can reduce reliability hardware and software interaction as 
\begin{equation}
    R_{HW\& SW}(t) = \frac{\lambda_{2}e^{-\lambda_{1}t}-\lambda_{1}e^{-\lambda_{2}t}}{\lambda_{2}-\lambda_{1}}.
\end{equation}
The estimated value of $(\lambda_{1},\lambda_{2})$ can be calculated by a non-linear sum of square estimates 
\begin{equation}
    SSE(\lambda_{1},\lambda_{2}) = \sum_{i=1}^{8}[X_{i}-\frac{T_{i}}{\frac{1}{\lambda_{1}}+\frac{1}{\lambda_{2}}}], 
\end{equation}
where $X_{i}$ is the aggregate exposure number, and $T_{i}$ is the aggregate exposure time.  As $(\lambda_{1},\lambda_{2}$ is not possible to estimate uniquely. Thus, we can rewrite$SSE(\lambda_{1},\lambda_{2})$ as $SSE(G,\lambda_{2})$, and $G=\lambda_{2}/\lambda_{1}$. Therefore the least-square estimates of $\lambda_{1}$ is 
\begin{equation}
    \lambda_{1} = \frac{(1+1/G)(\sum_{i=1}^8X_{i}T_{i})}{\sum_{i=1}^8 T^2},
\end{equation}
where $\lambda_{2}= G\lambda_{1}$.

\subsection{Computation Simulation Result }
The MATLAB simulation is given in the Fig.5. If we choose $G=2$, we can solve $\lambda_{1}=8.92\times10^{-4}$,$\lambda_{2}=3.92\times10^{-3}$.
\section{Conclusion}
In this paper, we first proposed unified hardware and software PMU. By applying fuzzy approach, we determine the reliability parameters. We use MATLAB to run simulation to do the parameter fitting, and finally we solve failure of hardware and software interaction failure.

\begin{figure*}[t]
  \includegraphics[width=2\columnwidth]{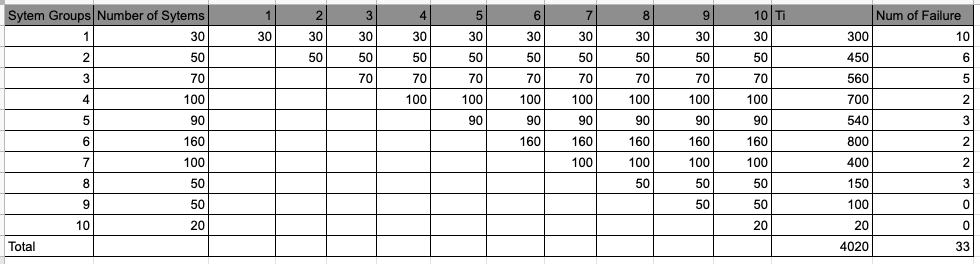}\caption{}
\end{figure*}









\section{Reference}
[1] C. Murthy, K. A. Varma, D. S. Roy, and D. K. Mohanta, “Reliability Evaluation of Phasor Measurement Unit Using Type-2 Fuzzy Set Theory,” IEEE Systems Journal, vol. 8, no. 4, pp. 1302–1309, Dec. 2014.

[2]'IEEE standard for synchrophasors for power system'. IEEE C37.118, 2005.

[3]Zhang, P., Chan, K.W.: ‘Reliability’, IEEE Trans. Smart Grid, 2012, 3, (3), pp. 1235–1243.

[4]Iyer, R.K., Velardi, P.: ‘Hardware-related software errors: measurement and analysis’, IEEE Trans. Softw. Eng., 1985, 2, pp. 223–231.

[5]Wang, Y., Li, W., Lu, J.: ‘Reliability analysis of phasor measurement unit using hierarchical Markov modeling’,
Electr. Power Compon. Syst., 2009, 37, (5), pp. 517–532

[6]Peng, Z.: ‘Wide area monitoring system and its application on power system low-frequency oscillation suppression’. PhD dissertation, Department of Electrical Engineering, Hong Kong Polytechnic University, 2012.

[7] Murthy, C., Mishra, A., Ghosh, D., Sinha Roy, D., Mohanta, D.K.: ‘Reliability analysis of phasor measurement unit using hidden Markov model’, IEEE Syst. J., 2014, in press, DOI: 10.1109/ JSYST.2014.2314811

[8] Xu, Z. and Pierre, J.W., 2022, October. Cramér-Rao Lower Bound for Forced Oscillations under Independent Multi-channel PMU measurements in Power Systems. In 2022 North American Power Symposium (NAPS) (pp. 1-6). IEEE.

[9]Xu, Z. and Pierre, J.W., 2021, November. Initial results for cramér-rao lower bound for forced oscillations in power systems. In 2021 North American Power Symposium (NAPS) (pp. 1-5). IEEE.

[10]Xu, Z., Soltani, M. and Singh, A., 2018, December. Exact statistical moments of multi-mode stochastic hybrid systems with renewal transitions. In 2018 IEEE Conference on Decision and Control (CDC) (pp. 3510-3515). IEEE.

[11]Xu, Z., Pierre, J.W., Elliott, R., Schoenwald, D., Wilches-Bernal, F. and Pierre, B., 2022, June. Cramér-Rao Lower Bound for Forced Oscillations under Multi-channel Power Systems Measurements. In 2022 17th International Conference on Probabilistic Methods Applied to Power Systems (PMAPS) (pp. 1-6). IEEE.

[12]Xu, Z., 2019. Analysis and applications of time-triggered stochastic hybrid system. University of Delaware.

\end{document}